\documentclass[11pt,a4paper]{article}

\usepackage{jheppub}
\usepackage[english]{babel}
\usepackage{amssymb,amsmath}
\usepackage{hyperref}
\usepackage{graphicx}
\usepackage{mathtools}

\newcommand{\vev}{{\langle \chi \rangle}}

\title{Forbidden conformal dark matter at a GeV}

\author[a]{Steven Ferrante,}
\author[a]{Ameen Ismail,}
\author[b]{Seung J. Lee,}
\author[b]{and Yunha Lee}
 
\affiliation[a]{Laboratory for Elementary Particle Physics, Cornell University, Ithaca, NY 14853, USA}
\affiliation[b]{Department of Physics, Korea University, Seoul 136-713, Korea}

\emailAdd{sef87@cornell.edu}
\emailAdd{ai279@cornell.edu}
\emailAdd{sjjlee@korea.ac.kr}
\emailAdd{dbsgk41@korea.ac.kr}

\abstract{We introduce a model of dark matter (DM) where the DM is a composite of a spontaneously broken conformal field theory. The DM is a thermal relic with its abundance determined by the freeze-out of annihilations to dilatons, the Goldstone boson of broken conformal symmetry. If the dilaton is heavier than the DM this is an example of forbidden DM. We explore the phenomenology of this model in its 5D dual description, corresponding to a warped extra dimension with the Standard Model on the ultraviolet brane and the DM on the infrared brane. We find the model is compatible with theoretical and experimental constraints for DM masses in the $0.1$--$10$~GeV range. The conformal phase transition is supercooled and strongly first-order. It can source large stochastic gravitational wave signals consistent with those recently observed at pulsar timing arrays like NANOGrav.
The majority of the viable parameter space will be probed by future detectors designed to search for long-lived particles, including most of the region favored by the NANOGrav signal.
The rest of the parameter space can be probed at future direct detection experiments.
}

\begin{document}

\maketitle	
\flushbottom

%%%%%%%%%%%%%%%%%%%%%%%%%%%%%%%%%%%%%%%%%%%%%%%%%%%%%%%%%%%
%%%%%%%%%%%%%%%%%%%%%%%%%%%%%%%%%%%%%%%%%%%%%%%%%%%%%%%%%%%
\section{Introduction}
%%%%%%%%%%%%%%%%%%%%%%%%%%%%%%%%%%%%%%%%%%%%%%%%%%%%%%%%%%%
%%%%%%%%%%%%%%%%%%%%%%%%%%%%%%%%%%%%%%%%%%%%%%%%%%%%%%%%%%%
Understanding the microscopic nature of dark matter (DM) is a paramount goal of elementary particle physics. The evidence for DM provides a strong experimental indication of new physics beyond the Standard Model (SM). The best-studied DM scenario is probably the weakly interacting massive particle (WIMP) paradigm, involving DM near the weak scale whose relic abundance is determined by the thermal freeze-out of annihilations to SM particles. This has a good theoretical motivation in that WIMPs arise naturally in many solutions to the Higgs hierarchy problem, and that weak-scale masses and cross sections coincidentally yield the observed DM relic abundance (the WIMP miracle). However, WIMP models face increasing pressure from the null results of direct detection experiments, motivating the exploration of other DM frameworks. In particular, the past 15 years have seen the emergence of a plethora of models in which the DM is a thermal relic, but much lighter than a typical WIMP~\cite{Kaplan:2009ag,Boddy:2014yra,Hochberg:2014dra,DAgnolo:2015ujb,Kuflik:2015isi,Pappadopulo:2016pkp,Dror:2016rxc,DAgnolo:2017dbv,Berlin:2017ftj,DAgnolo:2018wcn,Fitzpatrick:2020vba,Frumkin:2021zng}.

In this paper we introduce a model of thermal GeV-scale DM from a dark sector with spontaneously broken conformal symmetry. The DM is a composite of the conformal sector and the SM fields are taken to be elementary. Interactions between the dark and visible sectors are mediated by the dilaton, the Goldstone boson of spontaneously broken scale invariance. The DM relic abundance is set by $2 \rightarrow 2$ annihilations to dilatons. We will see that to achieve the observed relic abundance $\Omega_{\rm DM} h^2 \sim 0.1$, the DM must be somewhat lighter than the dilaton, making this model an example of forbidden dark mattter~\cite{DAgnolo:2015ujb}.

The AdS/CFT correspondence relates spontaneously broken conformal field theories (CFTs) to 5D warped setups with UV and IR branes~\cite{Rattazzi:2000hs,Arkani-Hamed:2000ijo}. The location of the UV brane corresponds to the cutoff of the CFT, while the IR brane is identified with the radion degree of freedom, which is dual to the dilaton~\cite{Csaki:1999mp,Csaki:2000zn}. Elementary states are localized towards the UV brane and composites of the CFT are localized towards the IR brane. Thus, the 5D dual description of our model involves a warped extra dimension where the DM propagates on the IR brane and the SM fields propagate on the UV brane.

In the context of physics beyond the SM, spontaneously broken CFTs are probably best known for their role in solutions to the Higgs hierarchy problem based on compositeness and their 5D duals~\cite{Randall:1999ee,Contino:2003ve,Agashe:2004rs,Bellazzini:2014yua,Panico:2015jxa} (as well as in some cosmological naturalness models~\cite{Csaki:2020zqz,Csaki:2022zbc}). The conformal symmetry stabilizes a large UV/IR hierarchy. The lightest Kaluza--Klein (KK) particle in these models can be a WIMP DM candidate, stabilized by KK-parity~\cite{Servant:2002aq,Cheng:2002ej,Agashe:2007jb,Medina:2011qc}, or by a $Z_3$ symmetry resulting from a requirement of proton stability~\cite{Agashe:2004ci, Agashe:2004bm}. There are numerous other DM models based on conformal sectors which go beyond the canonical paradigm of WIMP DM in a composite Higgs setting. These include conformal freeze-in~\cite{Hong:2019nwd,Hong:2022gzo,Chiu:2022bni,Redi:2021ipn}, continuum dark matter~\cite{Csaki:2021gfm,Csaki:2021xpy,Csaki:2022lnq,Ferrante:2023fpx} and other models featuring continua~\cite{Katz:2015zba,Chaffey:2021tmj,Fichet:2022ixi,Fichet:2022xol}, freeze-in from warped extra dimensions~\cite{Kolb:2003mm,Bernal:2020fvw,Bernal:2020yqg,deGiorgi:2021xvm,Cai:2021nmk,deGiorgi:2022yha}, dilaton-mediated models~\cite{Bai:2009ms,Agashe:2009ja,Blum:2014jca,Kim:2016jbz, Fuks:2020tam}, and others~\cite{Brax:2019koq,Redi:2020ffc,Ahmed:2023vdb,Ismail:2023afl}. Some of the freeze-in models accommodate light DM candidates, but to our knowledge, our model is the first example of light conformal DM with a thermal dark sector\footnote{The model in~\cite{Ahmed:2023vdb} can accommodate light DM in a small region of parameter space, but the authors primarily focus on masses heavier than $\sim 20$~GeV.}.

We first motivate our model in Section~\ref{sec:model}, arguing that the simplest way to achieve light, thermal, conformal DM is through the forbidden DM scenario. We realize our model through an extra-dimensional construction and compute the DM relic abundance. We also sketch how the model can easily be UV-completed into a composite Higgs model, although the DM physics is independent of the UV completion.

At high temperatures, we expect that the conformal sector is in its hot, deconfined phase with unbroken conformal symmetry. As the universe cools, a first-order phase transition takes place in which conformal symmetry is spontaneously broken. In Section~\ref{sec:pt} we investigate the conformal phase transition in the context of our model, studying under what conditions the phase transition is able to complete.

A complementary motivation for studying the phase transition is the recent observation of a nanoHertz (nHz)-scale stochastic gravitational wave background at multiple pulsar timing arrays~\cite{NANOGrav:2023gor,Xu:2023wog,Antoniadis:2023bjw,Reardon:2023gzh}. Such a signal has an astrophysical explanation, arising from inspiraling supermassive black hole binaries~\cite{NANOGrav:2023hfp}, although it can also originate from various new physics scenarios. The NANOGrav collaboration released an analysis of their signal in the context of a handful of well-motivated frameworks for new physics~\cite{NANOGrav:2023hvm}, finding that cosmic inflation~\cite{Guzzetti:2016mkm}, scalar-induced gravitational waves~\cite{Domenech:2021ztg,Yuan:2021qgz}, first-order phase transitions~\cite{Caprini:2015zlo,Caprini:2019egz}, and topological defects~\cite{Vilenkin:1984ib,Hindmarsh:1994re,Saikawa:2017hiv} are all consistent with their results. Curiously, these models (with or without a contribution from supermassive black hole binaries included) actually provide somewhat better fits to the NANOGrav data than supermassive black hole binaries alone. The conformal phase transition has recently attracted renewed attention as a possible source of the NANOGrav signal~\cite{Fujikura:2023lkn,Megias:2023kiy}. For this reason, in Section~\ref{sec:pt} we also examine the gravitational wave signals from the phase transition in our model. We find a gravitational wave signal consistent with the NANOGrav result for dilaton masses ranging from $0.1$--$3$~GeV.

Section~\ref{sec:pheno} focuses on the phenomenology of our model. The requirement that the dark sector is thermalized with the SM after the conformal phase transition and bounds from direct detection experiments constrain the dilaton and DM masses to lie around $0.1$--$10$~GeV. Much of the viable parameter space for masses below $\sim 5$~GeV, including that favored by the NANOGrav signal, will be probed in the near future by experiments searching for light, weakly-coupled particles. In a complementary fashion, the remaining parameter space in the $5$--$10$~GeV mass range can be explored at future direct detection experiments.

%%%%%%%%%%%%%%%%%%%%%%%%%%%%%%%%%%%%%%%%%%%%%%%%%%%%%%%%%%%
%%%%%%%%%%%%%%%%%%%%%%%%%%%%%%%%%%%%%%%%%%%%%%%%%%%%%%%%%%%
\section{Through the dilaton portal}\label{sec:model}
%%%%%%%%%%%%%%%%%%%%%%%%%%%%%%%%%%%%%%%%%%%%%%%%%%%%%%%%%%%
%%%%%%%%%%%%%%%%%%%%%%%%%%%%%%%%%%%%%%%%%%%%%%%%%%%%%%%%%%%

\subsection{Why forbidden DM?}

Before introducing our detailed model, it is instructive to consider DM from a spontaneously broken conformal sector more generally. We will see that the requirements that the DM lies near the GeV scale and that the dark sector is thermalized with the SM naturally lead us to forbidden DM.

The dark sector must contain a dilaton field $\sigma$, the Goldstone boson of broken scale invariance, so one might minimally consider a model where the dilaton is the DM. However, one expects the dilaton to have couplings to the light SM fermions, suppressed by the cutoff scale of the CFT. This typically allows the dilaton to decay to $e^+ e^-$ pairs, which rules out the dilaton as the DM unless its lifetime is larger than about $10^{25}$~s~\cite{Slatyer:2016qyl}. It is possible to avoid this issue by taking the cutoff of the CFT sufficiently high, but then the interactions between the SM and the dark sector are too feeble to keep them in thermal contact. In that case one could still generate the DM relic abundance by a nonthermal mechanism, such as conformal freeze-in. Note that in this case the dark sector would need to be cooler than the SM sector to obtain the correct relic abundance, which severely damps the gravitational wave signals from the conformal phase transition~\cite{Breitbach:2018ddu,Ertas:2021xeh,Bringmann:2023opz}. An observable stochastic gravitational wave background is only generated if the dark sector temperature is comparable to or larger than the visible sector temperature.

This motivates the introduction of a composite field $\phi$ (in the 5D picture, a field localized on the IR brane) to be the DM.
What mechanism can set the relic abundance of $\phi$? The simplest option is for $\phi$ to be a canonical WIMP that freezes out through $2 \rightarrow 2$, dilaton-mediated annihilations to SM particles. At $T \lesssim m_\phi$, the thermally averaged cross section for this process is at most of order $\langle \sigma v \rangle \sim m_\phi^2 / \Lambda^4$, where $\Lambda$ is the cutoff of the CFT. For $m_\phi \sim $~GeV and $\Lambda \sim $~TeV, we obtain $\langle \sigma v \rangle \sim (10^3 {\rm~TeV})^{-2}$, which is much smaller than the value that would reproduce the observed DM relic abundance, namely $(20 {\rm~TeV})^{-2}$.
One might attempt to solve this issue by reducing the cutoff to $\mathcal{O}(100{\rm ~GeV})$, but this would result in a large DM-nucleon cross section in tension with direct detection bounds. (We will further discuss the experimental bounds on our model in section~\ref{sec:pheno}.

We conclude that $\phi$ cannot be a minimal WIMP because its interactions with the SM are too weak. However, its interactions with the dilaton are suppressed by the symmetry-breaking IR scale $f$, not by $\Lambda$, so the freeze-out of $\phi$ may be controlled by annihilations to dilatons instead. If $m_\phi > m_\sigma$ this is a secluded DM scenario~\cite{Pospelov:2007mp}, while if $m_\phi < m_\sigma$ it is a forbidden DM scenario~\cite{DAgnolo:2015ujb} (see~\cite{Delgado:2016umt,Aoki:2016glu,DAgnolo:2020mpt,Wojcik:2021xki,Yang:2022zlh,Cheng:2023hzw} for other forbidden DM models). In the secluded case, we can estimate the $\phi\phi \rightarrow \sigma\sigma$ annihilation cross section at $T \lesssim m_\phi$ as $1 / m_\phi^2$. For GeV-scale dark matter this is too large to obtain the right relic abundance.

This leads us to the forbidden DM case. GeV-scale dark matter is viable in this case because the annihilation cross section is exponentially suppressed by Boltzmann factors. To summarize, we assume that conformal symmetry is broken at a scale $m_\sigma \sim f \sim $~GeV and that $m_\phi \lesssim m_\sigma$. The process governing the freeze-out of the $\phi$ is $2 \rightarrow 2$ annihilations to dilatons, $\phi \phi \rightarrow \sigma \sigma$.

\subsection{5D model}

\begin{figure}
\centering
\includegraphics[width=6in]{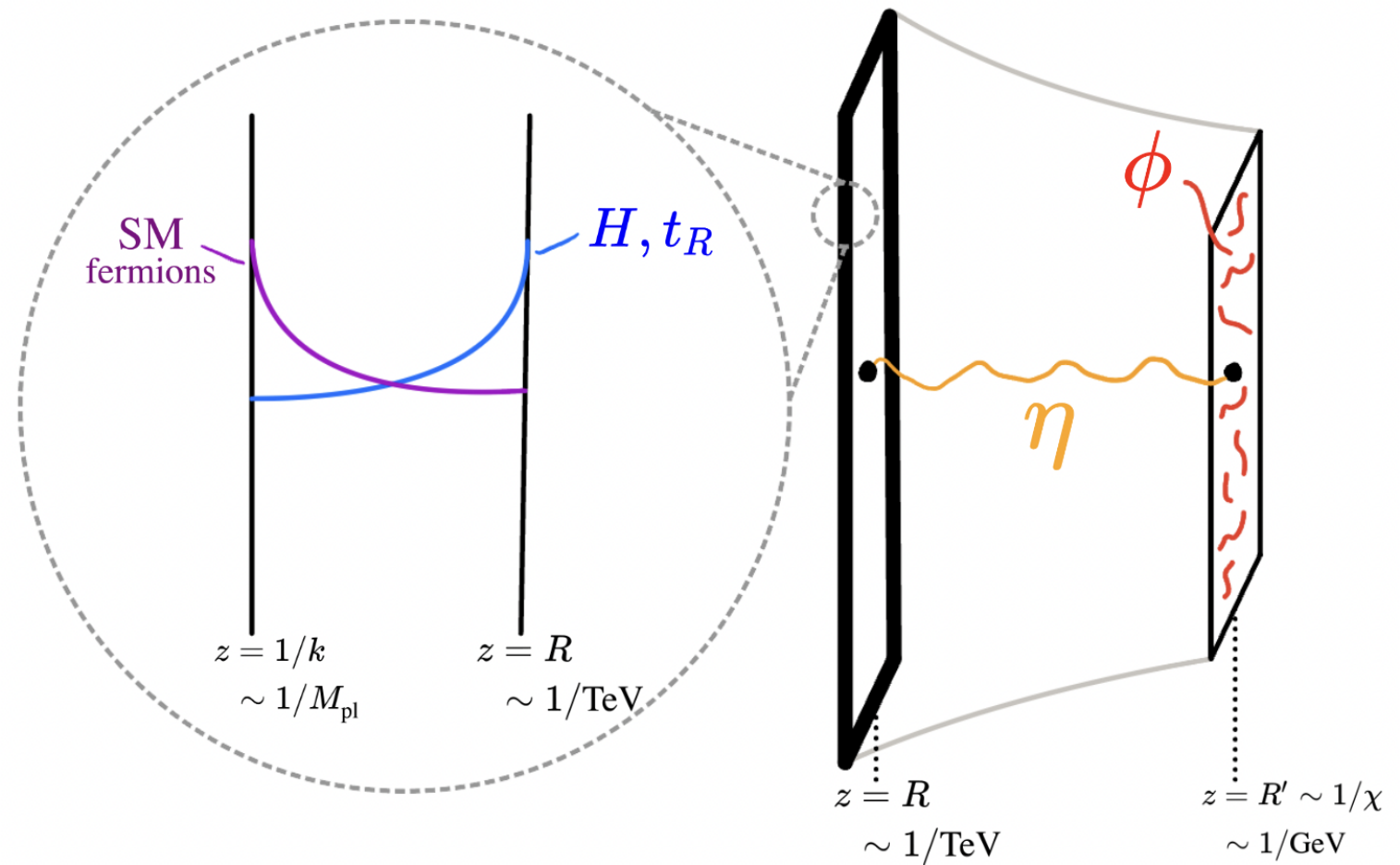}
    \caption{5D setup of forbidden conformal DM (see text for details). The Goldberger--Wise stabilization field $\eta$ propagates in the bulk of a warped extra dimension, the DM field $\phi$ propagates on the GeV-scale IR brane, and the SM fields propagate on a TeV-scale UV brane. The inset depicts how this setup may be UV-completed into a typical composite Higgs model, where the TeV brane is an intermediate brane and there is a true UV brane at a much higher scale.}
\label{fig:5dSketch}
\end{figure}

We now realize this picture of forbidden composite DM in the 5D warped description of the CFT~\cite{Randall:1999ee}. Consider a slice of 5D AdS space with curvature $k$, described by the metric\footnote{We use Greek indices for 4D coordinates and Latin indices for 5D coordinates. 4D (5D) coordinates are raised and lowered with the Minkowski metric $\eta_{\mu\nu}$ (the full metric $g_{MN}$).}
\begin{equation}\label{eq:5Dmetric}
    ds^2 = \frac{1}{k^2z^2} \left( \eta_{\mu\nu} dx^\mu dx^\nu - dz^2 \right) ,
\end{equation}
where the extra dimension extends from a UV brane at $z = R \sim 1/{\rm TeV}$ to an IR brane at $z = R' \sim 1/{\rm GeV}$, as shown in Fig.~\ref{fig:5dSketch}. This is a solution to the Einstein equations with bulk cosmological constant $\Lambda_{\rm CC} = -24 M_5^3 k^2$ and brane tensions $\pm \Lambda_{\rm CC}/k$, where $M_5$ is the 5D Planck scale. With the cosmological constant and brane tensions set to these values, the 5D Einstein--Hilbert action is
\begin{equation}
    S_{\rm EH} = \int d^5 x \sqrt{g} \left(-2 M_5^3 R - \Lambda_{\rm CC} \right) - \sqrt{\tilde{g}} \Lambda_{\rm CC} \frac{\delta(z - R)}{k} + \sqrt{\tilde{g}} \Lambda_{\rm CC} \frac{\delta(z-R')}{k}
\end{equation}
where $\tilde{g}$ denotes the induced metric.

We identify the dilaton field $\chi = 1/R'$ with the location of the IR brane~\cite{Csaki:1999mp,Csaki:2000zn,Rattazzi:2000hs}.
For simplicity, we take $\phi$ to be a real scalar field which propagates on the IR brane, and assume that $\phi$ is odd under a stabilizing $\mathbb{Z}_2$ symmetry. The action for $\phi$ is
\begin{equation}
    S_\phi = \int d^5 x \sqrt{\tilde{g}} \delta(z - R') \left[ \frac{1}{2} (\partial_\mu \phi)^2 - \frac{1}{2} m_\phi^2 \phi^2 - \frac{\lambda_\phi}{4!} \phi^4 \right] .
\end{equation}
The dark sector consists of $\chi$ and $\phi$, where $\phi$ will ultimately play the role of the DM. The quartic coupling $\lambda_\phi$ will only play an important role when we consider DM self-interactions and can otherwise be ignored.

Note that $R \gg 1/k$ in our model, unlike a typical RS model where the UV brane lies at $R = 1/k$ and corresponds to the cutoff scale of the dual CFT. We expect this setup could be UV-completed into a three-brane model (see Fig.~\ref{fig:5dSketch} and for examples see~\cite{Agashe:2016rle,Csaki:2016kqr,Lee:2021wau,Girmohanta:2023sjv}), where the true, Planck-scale UV brane lies at $z = 1/k$, and the SM fields propagate in the bulk between the Planck and TeV branes similar to a holographic composite Higgs or realistic RS model. In this case the light fermions would be mostly elementary and thus localized toward the Planck brane, while the right-handed top quark and Higgs would be localized toward the TeV brane. However, for the purposes of our DM model the details of the UV completion do not matter, and we simply take the SM fields to lie on the TeV brane.

Finally, we introduce a bulk real scalar field $\eta$ which will serve to stabilize the radius of the extra dimension (dual to the scale of conformal symmetry breaking) via the Goldberger--Wise mechanism~\cite{Goldberger:1999uk}. For simplicity we also assume $\eta$ is odd under an unbroken $\mathbb{Z}_2$ symmetry, leading to the action
\begin{equation}
    S_{\rm GW} = \int d^5 x \sqrt{g} \left[ \frac{1}{2} ( \partial_M \eta )^2 - \frac{1}{2} m_\eta^2 k^2 \eta^2 \right] - \sqrt{\tilde{g}} \delta(z - R) V_{\rm UV}(\eta) - \sqrt{\tilde{g}} \delta(z - R') V_{\rm IR}(\eta) .
\end{equation}
For the brane-localized potentials, we choose
\begin{equation}
    V_{\rm UV}(\eta) = \beta \left(\eta^2 - k^3 v_\eta^2 \right)^2 , \quad V_{\rm IR} = \frac{1}{2} k m_{\rm IR} \eta^2
\end{equation}
such that $\eta$ gets a vacuum expectation value (vev) on the UV brane.

\subsection{4D effective Lagrangian}
One can integrate out the bulk to obtain an effective 4D action for the model~\cite{Bellazzini:2013fga,Csaki:2022htl,Csaki:2023pwy}. The dilaton kinetic term arises from the 5D Einstein--Hilbert action, while the vev of the bulk field $\eta$ induces a potential for the dilaton $\propto \chi^{4+\alpha}$, where $\alpha = 2(\sqrt{4+ m_\eta^2} - 2)$. The effective dilaton action, not yet including interactions with brane-localized fields, takes the usual Goldberger--Wise form
\begin{equation}\label{eq:potentialRaw}
    \int d^4 x \frac{3N^2}{4\pi^2} \left( \partial_\mu \chi \right)^2 - V(\chi), \quad V(\chi) = \frac{3N^2}{2\pi^2} \left[ - \lambda \chi^4 + \lambda_{\rm GW} \frac{\chi^{4+\alpha}}{R^{-\alpha}} \right ] + V_0.
\end{equation}
Here $N$ is the number of colors in the dual CFT, related to the 5D parameters by $N^2 = 16\pi^2(M_5/k)^3$ (for example, see~\cite{vonHarling:2017yew}). The dilaton quartic $\lambda$ and constant term $V_0$ arise from a mistuning of the IR brane and UV brane tensions, respectively. The coefficient of the Goldberger--Wise term $\lambda_{\rm GW}$ is an order-one function of $v_\eta$, $m_{\rm IR}$, and $\alpha$.

The dilaton acquires a vev
\begin{equation}
    \vev = R^{-1} \left( \frac{\lambda}{\lambda_{\rm GW}} \right)^{1/\alpha} .
\end{equation}
For small $\alpha$ an $\mathcal{O}(1)$ ratio $\lambda/\lambda_{\rm GW} < 1$ creates a large UV-IR hierarchy, $\vev \ll R^{-1}$. For a modest TeV--GeV hierarchy a typical value of $\alpha$ is $1/\log (10^{3}) \sim 0.1$--$0.2$. We then reparametrize the dilaton potential in terms of its vev and tune the constant term so that $V(\vev) = 0$, corresponding to the vanishing of the 4D cosmological constant. This yields
\begin{equation}\label{eq:potential}
    V(\chi) = \frac{3N^2}{2\pi^2} \lambda \vev^4 \left[ 1 - \left( \chi/\vev  \right)^4 + \frac{-1 + \left( \chi/\vev \right)^{4+\alpha} }{1 + \alpha/4} \right] .
\end{equation}
The dilaton mass-squared is $4 \alpha \lambda \vev^2$.

The couplings of the dilaton to $\phi$ are fixed by scale invariance and can be computed with a simple spurion analysis~\cite{Goldberger:2007zk,Bai:2009ms,Blum:2014jca,Fuks:2020tam}. The only scale in the $\phi$ Lagrangian is its mass $m_\phi$. Thus we make the replacement $m_\phi \rightarrow m_\phi \cdot \chi / \vev$ to make the action for $\phi$ scale-invariant, resulting in
\begin{equation}
    \int d^4 x \frac{1}{2} (\partial_\mu \phi)^2 - \frac{\chi^2}{2 \vev^2} m_\phi^2 \phi^2 .
\end{equation}
Note that the mass of $\phi$ gets warped down, so the natural size of $m_\phi$ is determined by $\vev$ and not by $R^{-1}$. The leading-order couplings of the dilaton to the SM are obtained in a similar way: we replace each mass as $m \rightarrow m \cdot \chi / \vev \cdot (R \vev)^2$. The extra factor of $(R\vev)^2$ is the wavefunction suppression arising from the fact that the dilaton wavefunction is peaked on the IR brane~\cite{Charmousis:1999rg,Csaki:2000zn}.

Finally, we rewrite the Lagrangian in terms of canonically normalized fluctuations about the dilaton vev $\chi = \vev + \sqrt{2\pi^2/3N^2} \sigma$. We further eliminate the dilaton quartic $\lambda$ in favor of the dilaton mass $m_\sigma^2 = 4\alpha\lambda \vev^2$, define $f = \sqrt{3N^2/2\pi^2} \vev$ (the symmetry-breaking scale), and define $\Lambda = \sqrt{3N^2/2\pi^2}R^{-1}$ (the UV cutoff). Our final tree-level effective Lagrangian is then
\begin{equation}\label{eq:effL}\begin{split}
    \mathcal{L} &= \mathcal{L}_{\rm SM} + \frac{1}{2} (\partial_\mu \sigma)^2 - \frac{1}{2} m_\sigma^2 \sigma^2 - \frac{5}{6} \frac{m_\sigma^2}{f} \sigma^3 - \frac{11}{24} \frac{m_\sigma^2}{f^2} \sigma^4 \\
    &+ \frac{1}{2} (\partial_\mu \phi)^2 - \frac{1}{2} m_\phi^2 \phi^2  - \frac{1}{4!} \lambda_\phi^4  - \left( \frac{2 \sigma}{f} + \frac{\sigma^2}{f^2} \right) \frac{1}{2} m_\phi^2 \phi^2 \\
    & -\frac{\sigma}{\Lambda^2/f} \left[ \sum_{\rm fermions} m_\psi \overline{\psi} \psi + m_h^2 h^2 - 2 m_W^2 W_\mu^+ W^{-\mu} - m_Z^2 Z_\mu Z^\mu \right] .
\end{split}\end{equation}
In the above equation, the first line includes terms up to order $\chi^4$ in the dilaton potential and at leading order in $\alpha$; the second line is the Lagrangian for the DM $\phi$ and its couplings to the dilaton; and the last line is the interactions of the dilaton with SM fields, where the sum runs over all SM fermions. We have only written the dilaton-SM interactions at leading order in $1/\Lambda$.

Due to the trace anomaly, the dilaton further acquires couplings to the gluon (photon) proportional to the QCD (QED) $\beta$-function and couplings to fermions proportional to their anomalous dimensions $\gamma_\psi$~\cite{Damour:2010rp,Kaplan:2000hh,Csaki:2007ns}. These take the form
\begin{equation}
    -\frac{\sigma}{\Lambda^2/f} \left[ \frac{\beta_e (e)}{2 e^3} F_{\mu\nu}^2 + \frac{\beta_3 (g_3)}{2 g_3^3} \left(G^a_{\mu\nu}\right)^2 + \sum_{\rm fermions} \gamma_\psi \overline{\psi} \psi \right] ,
\end{equation}
where $\beta_e$ ($\beta_3$) is the $\beta$-function for the electromagnetic (QCD) coupling $e$ ($g_3$). The net effect of the gluon and quark couplings is that the dilaton effectively couples to the total mass of hadrons. The photon coupling is only phenomenologically important when the dilaton is lighter than $\sim 1$~MeV and $\sigma \rightarrow \gamma \gamma$ is the only accessible decay mode.

\subsection{Relic abundance}

Armed with the effective Lagrangian in Eq.~\eqref{eq:effL}, we can compute the relic abundance of $\phi$. The Boltzmann equations for the DM $\phi$ and the dilaton $\sigma$ are given by
\begin{equation}\label{eq:boltzmann}
    \begin{split}
        \dot{n}_\phi + 3 H n_\phi &= n_\sigma^2 \langle \sigma v (\sigma\sigma \rightarrow \phi\phi) \rangle - n_\phi^2 \langle \sigma v (\phi\phi \rightarrow \sigma\sigma) \rangle ,  \\
        \dot{n}_\sigma + 3 H n_\sigma &= n_\phi^2 \langle \sigma v (\phi\phi \rightarrow \sigma\sigma) \rangle - n_\sigma^2 \langle \sigma v (\sigma\sigma \rightarrow \phi\phi) \rangle + {\rm SM~interactions} .
    \end{split}
\end{equation}
In the second line, the SM interactions refer to both $2\rightarrow 2$ processes of the form $\sigma \sigma \leftrightarrow f \overline{f}$, where $f$ is any light SM fermion, and decays and inverse decays $\sigma \leftrightarrow f \overline{f}$. These interactions are responsible for keeping the dilaton in thermal equilibrium with the SM. As a first approximation, we assume that the dilaton remains in equilibrium while $\phi$ freezes out. Under this assumption we have a straightforward forbidden DM model~\cite{DAgnolo:2015ujb}. (In practice, the dilaton may freeze out somewhat before the  DM, so we will numerically solve the Boltzmann equations to obtain more precise results for the relic abundance.)

Using Eq.~\eqref{eq:effL}, we find the thermally averaged s-wave annihilation cross section for $\sigma\sigma \rightarrow \phi\phi$ in the nonrelativistic limit:
\begin{equation}\label{eq:swave}
    \langle \sigma v(\sigma\sigma \rightarrow \phi\phi) \rangle  = \frac{1}{9\pi m_\phi^2} \left( \frac{m_\phi}{f} \right)^4 \frac{\sqrt{\Delta(2+\Delta)}}{(1+\Delta)^7} \left(1 - 4\Delta - 2\Delta^2 \right)^2 
\end{equation}
where $\Delta = (m_\sigma - m_\phi)/m_\phi$ is the relative mass splitting. Note that we have ignored p-wave and higher-order contributions, which are only important for $\Delta$ very close to $\sqrt{3/2} - 1 \approx 0.22$, where Eq.~\eqref{eq:swave} vanishes. The cross section for the reverse process follows from the principle of detailed balance~\cite{DAgnolo:2015ujb},
\begin{equation}\begin{split}
    \langle \sigma v(\phi\phi \rightarrow \sigma\sigma) \rangle &= \left( \frac{n_\sigma^{\rm eq}}{n_\phi^{\rm eq}} \right)^2 \langle \sigma v(\sigma\sigma \rightarrow \phi\phi) \rangle  \\
    &= \frac{1}{9\pi m_\phi^2} \left( \frac{m_\phi}{f} \right)^4 \frac{\sqrt{\Delta(2+\Delta)}}{(1+\Delta)^4} \left(1 - 4\Delta - 2\Delta^2 \right)^2 e^{-2\Delta x} ,
\end{split}\end{equation}
with $x = m_\phi/T$.

Following~\cite{DAgnolo:2015ujb}, the DM relic abundance is given by
\begin{equation}\label{eq:relicabundance}\begin{split}
    \Omega_{\phi} h^2 &\sim 0.1 g_\Delta(x_f) \frac{9 \pi (f/m_\phi)^4 m_\phi^2}{(20 {\rm~TeV})^2} e^{2 \Delta x_f} , \\
    g_\Delta(x_f) &= \frac{2 (1+\Delta)^4}{\sqrt{\Delta(2+\Delta)} \left(1 - 4\Delta - 2\Delta^2 \right)^2} \left[ 1 - 2 \Delta x_f e^{2 \Delta x_f} \int_{2\Delta x_f}^\infty dt \frac{e^{-t}}{t} \right]^{-1} .
\end{split}\end{equation}
In the above equation, $x_f \sim 20$ is the value of $x$ at which the DM freezes out. One solves for $x_f$ by determining the temperature at which $n_\phi^{\rm eq} \langle \sigma v(\phi\phi \rightarrow \sigma\sigma) \rangle = H$. We reiterate that this computation of the relic abundance makes the simplifying assumption that the dilaton is in thermal equilibrium with the SM; to obtain a precise result one must numerically solve the Boltzmann equation.

%%%%%%%%%%%%%%%%%%%%%%%%%%%%%%%%%%%%%%%%%%%%%%%%%%%%%
\begin{figure}
\centering
\includegraphics[width=6in]{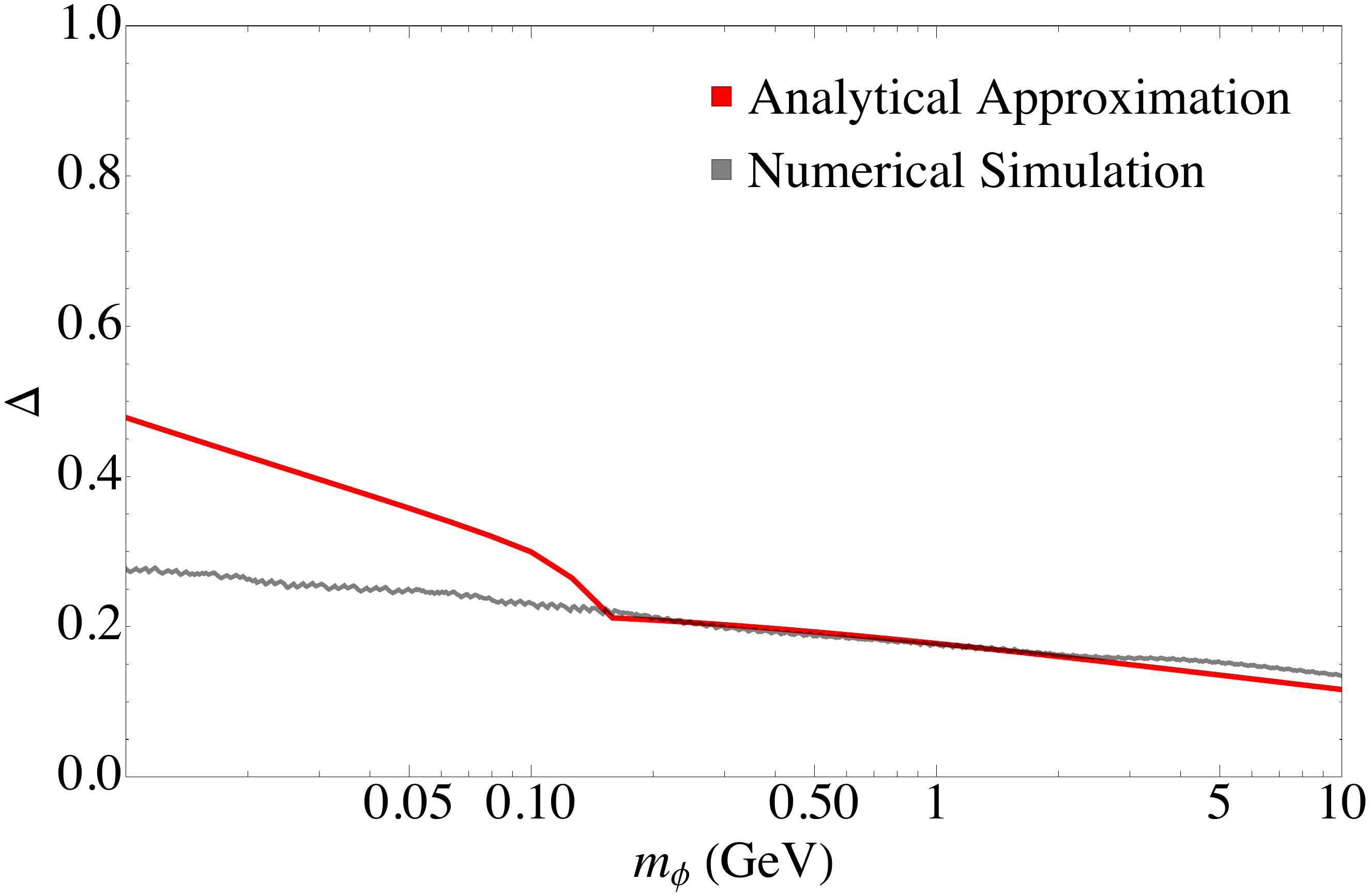}
    \caption{The mass splitting $\Delta = m_\sigma/m_\phi - 1$ which yields the observed DM relic abundance as a function of the DM mass $m_\phi$, fixing $f = m_\sigma$ and $\Lambda = 5$~TeV. We show the analytical approximation Eq.~\eqref{eq:relicabundance} \textit{(red line)} as well as numerical results computed using \texttt{micrOMEGAs} \textit{(gray line)}.}
\label{fig:relicabundance}
\end{figure}
%%%%%%%%%%%%%%%%%%%%%%%%%%%%%%%%%%%%%%%%%%%%%%%%%%%%%

In Fig.~\ref{fig:relicabundance} we show a contour in the $(m_\phi, \Delta)$ plane that yields the observed relic abudance $\Omega_{\phi} h^2 \sim 0.1$~\cite{Planck:2015fie}, fixing $\Lambda = 5$~TeV and $f = m_\sigma$. We include both the analytical approximation, Eq.~\eqref{eq:relicabundance}, and numerical results computed using \texttt{micrOMEGAs}~\cite{Belanger:2018ccd}. For $m_\phi \gtrsim 0.2$~GeV there is good agreement between the numerical and analytical results. We have not yet fully understood the origin of the discrepancy at smaller masses, but we suspect it is due to the dilaton falling out of thermal equilibrium before the DM, which renders the analytical approximation in Eq.~\eqref{eq:relicabundance} invalid.

%%%%%%%%%%%%%%%%%%%%%%%%%%%%%%%%%%%%%%%%%%%%%%%%%%%%%%%%%%%
%%%%%%%%%%%%%%%%%%%%%%%%%%%%%%%%%%%%%%%%%%%%%%%%%%%%%%%%%%%
\section{Conformal phase transition}\label{sec:pt}
%%%%%%%%%%%%%%%%%%%%%%%%%%%%%%%%%%%%%%%%%%%%%%%%%%%%%%%%%%%
%%%%%%%%%%%%%%%%%%%%%%%%%%%%%%%%%%%%%%%%%%%%%%%%%%%%%%%%%%%

At high temperatures we expect the conformal sector to be in its hot, deconfined phase with unbroken conformal symmetry. In the 5D picture this corresponds to a geometry with a UV brane and a black brane horizon instead of an IR brane. (When the UV brane is taken to the AdS boundary at $z = 0$, the hot phase geometry coincides with an AdS-Schwarzschild spacetime.) As the universe cools, we expect conformal symmetry to be broken as the CFT undergoes a phase transition to the cold, confined phase, which is dual to the 5D model described in the previous section. The phase transition proceeds via nucleation of bubbles of the IR brane. The conformal phase transition in holographic settings has been extensively studied in the literature~\cite{Creminelli:2001th,Randall:2006py,Nardini:2007me,Konstandin:2010cd,Konstandin:2011dr,Bunk:2017fic,vonHarling:2017yew,Pomarol:2019aae,Fujikura:2019oyi,Megias:2020vek,Bigazzi:2020phm,Agashe:2020lfz,Ares:2020lbt,Agrawal:2021alq,Levi:2022bzt,Csaki:2023pwy,Eroncel:2023uqf}, so here we just summarize the essential aspects.

It is important to study the phase transition in our model for two reasons. First, we must check that the phase transition is actually able to complete --- otherwise the conformal sector remains in the hot phase and there is no DM candidate. Second, bubble collisions during the phase transition can source stochastic gravitational waves, and we are interested in whether the phase transition could yield a gravitational wave signal consistent with that recently seen at the NANOGrav experiment.

\subsection{Phase transition completion}

Before checking whether the phase transition can complete, we first extract the critical temperature $T_{c}$ as a function of our model parameters by equating the free energies of the hot and cold phases. The free energy of the conformal sector in the cold phase is approximately (assuming a low critical temperature) given by $F_{\rm confined}(\chi) = V(\chi)$, with $V(\chi)$ defined in Eqs.~\eqref{eq:potentialRaw}--\eqref{eq:potential}. Note that this vanishes at $\chi = \langle \chi \rangle$ due to the tuning of the constant in the dilaton potential $V_0$. In the hot phase, the free energy is $F_{\rm deconfined}(T) = V_{0} - \pi^2 N^{2}T^{4}/8$~\cite{Creminelli:2001th}. We can then identify the critical temperature:
\begin{align}\label{eq:crit}
    F_{\text{confined}}(\langle \chi \rangle) = F_{\text{deconfined}}(T_c)
    \implies 
    T_{c} = 
    \sqrt{\frac{m_{\sigma} f}{\pi N}}
    \bigg(
    \frac{2}{4+\alpha}
    \bigg)^{1/4} .
    % \frac{\langle\sigma\rangle}{\pi}
    % \bigg(
    % \frac{12\epsilon\lambda}{4+\epsilon}
    % \bigg)^{1/4}
\end{align}
Note that in deriving $T_c$ we have neglected the contribution of $\phi$ to the free energy of the confined phase. Including this effect would give a $1/N^2$-suppressed correction to $T_c$.

To check whether the phase transition completes (and find the nucleation temperature $T_{n}$ at which it does so), we require that the probability of bubble nucleation per unit volume per unit time $\Gamma$ is greater than the Hubble parameter $H^{4}$. This probability can be approximated as $\Gamma \sim T_{n}^{4}e^{-S_{b}}$, where $S_{b}$ is the Euclidean bounce action. As long as $T_n/T_c$ is not very close to $1$, we have $T_{n}^{4}\ll T_{c}^{4}$, meaning the vacuum energy of the CFT dominates over the energy of the radiation bath before the phase transition: $\rho \approx \pi^2 N^2 T_c^4 / 8$. Therefore the Hubble parameter is approximately $H \sim \sqrt{\rho}/M_{\rm Pl} \sim T_{c}^{2}/M_{\rm Pl}$, and we find that the phase transition only completes if~\cite{Agashe:2019lhy,vonHarling:2017yew}
\begin{align}\label{eq:PTcompletion}
    S_{b} \lesssim
            4 \bigg( 
            \text{log}\,\frac{M_{\rm Pl}}{T_{c}}
            + 
            \text{log}\,\frac{T_{n}}{T_{c}}
            \bigg). 
\end{align}
For a GeV-scale $T_c$ the phase transition does not complete until $S_b \lesssim 170 + 4 \log(T_n / T_c)$.

The $O(3)$-symmmetric bounce action in our model $S_{b} = S_{3}/T_{n}$ can be calculated in the thick-wall limit, which is a good approximation for a supercooled transition~\cite{Linde:1981zj,vonHarling:2017yew}: 
\begin{align}
    S_{3}\approx 
    \frac{\sqrt{3}}{\pi^{2}}
    \frac{N^{3}\chi_{r}^{3}}
    {\sqrt{
    V(\langle\chi\rangle)(T_{n}/T_{c})^{4} - V(\chi_{r})
    }},
\end{align}
where $\chi_{r}$ is the ``release point'' of the phase transition (the value of the dilaton field that minimizes $S_{3}$). 
The nucleation temperature $T_{n}$ is found as the point at which the inequality Eq.~\eqref{eq:PTcompletion} is saturated.

\subsection{Gravitational wave signal}

If the phase transition can complete, we can describe the stochastic gravitational waves that it sources. Two characteristic features of the signal are the peak fractional abundance and the peak frequency of the gravitational waves. Assuming the signal is dominated by bubble wall collisions and the bubble wall velocity is $1$, these are given by~\cite{Caprini:2015zlo,Caprini:2019egz}
\begin{align}\label{eq:GWparameters}
    \Omega_{\text{GW}}h^{2} &\approx
    1.3 \times 10^{-6} 
    \bigg(
    \frac{H}{\beta_{\text{GW}}}
    \bigg)^{2}
    \bigg(
    \frac{100}{g_{*}}
    \bigg)^{1/3}
    \nonumber\\
    f_{\text{GW}} &\approx
    0.04 \,\, \text{mHz}
    \bigg(
    \frac{\beta_{\text{GW}}}{H}
    \bigg)
    \frac{T_R}{\text{TeV}}
    \bigg(
    \frac{g_{*}}{100}
    \bigg)^{1/6},
\end{align}
which depend on the duration $1/\beta_{GW}$ and the reheating temperature $T_R$ of the phase transition. (In the case of a phase transition without substantial reheating, one would use $T_n$ instead of $T_R$.) For a phase transition dominated by bubble collisions and including a contribution from supermassive black hole binaries, referred to as \texttt{PT-BUBBLE+SMBHB} in~\cite{NANOGrav:2023hvm}, the NANOGrav data favor $T_R \in (0.017,3.3)$~GeV and $\beta_{\rm GW}/H < 27$ at the 95\% credible level~\cite{NANOGrav:2023hvm}\footnote{We neglect correlations between $\beta_{\rm GW}/H$ and $T_R$ in this work.}. Note that in Eq.~\eqref{eq:GWparameters}, $g_*(T_R)$ counts the number of relativistic degrees of freedom in both the SM and dark sectors.

The phase transition duration can be extracted from the bounce action via~\cite{Caprini:2015zlo,Caprini:2019egz}
\begin{align}
    \frac{\beta_{GW}}{H} =
    T\frac{dS_{b}}{dT}
    \bigg|_{T_{n}}.
\end{align}
Free energy released during the phase transition is dumped into the SM and the conformal sector, reheating the universe. The reheating temperature $T_R$ can be calculated using energy conservation. Immediately after the phase transition the energy density is $\pi^{2}g_{*}(T_{R})T_{R}^{4}/30$. As previously mentioned, the vacuum energy $\pi^{2}N^{2}T_{c}^{4}/8$ is dominant before the transition, so the reheating temperature is
\begin{align}
    T_{R}^{4} = 
    \frac{15}{4}\frac{N^{2}}{g_{*}(T_{R})}
    T_{c}^{4}
    =  \frac{15}{2\pi^2(4+\alpha)} \frac{f^2 m_\sigma^2}{g_*(T_R)}
\end{align} 
using Eq.~\eqref{eq:crit} in the last equality. For a GeV-scale phase transition $g_*(T_R) \approx 70$, so $T_R \approx 0.2 \sqrt{m_\sigma f}$.

It is also important to ensure we have a strong phase transition in our model, which is characterized by the ratio of the energy released during the phase transition to the energy of the radiation bath~\cite{Caprini:2015zlo,Caprini:2019egz}:
\begin{align}\label{eq:PTstrength}
    \alpha_{\rm GW} = \frac{15N^{2}}{4g_{*}(T_n)}
    \bigg(
    \frac{T_{c}^{4}}{T_{n}^{4}}-1
    \bigg) .
\end{align}
The gravitational wave amplitude is proportional to $\alpha_{\rm GW}^2/(\alpha_{\rm GW}+1)^2$, which results in a damping of the signal for a weak phase transition with $\alpha_{\rm GW} \ll 1$. Also, the assumptions of relativistic bubble walls and a signal dominated by bubble collisions are only valid in the large $\alpha_{\rm GW}$ limit. For a supercooled phase transition $T_c^4 \gg T_n^4$, Eq.~\eqref{eq:PTstrength} predicts a strong phase transition, $\alpha_{\rm GW} \gg 1$ (for a more careful analysis see~\cite{Csaki:2023pwy}). In the region of our parameter space where the NANOGrav signal can be reproduced, $T_n/T_c$ is at most $\mathcal{O}(1\%)$; therefore we expect a strong phase transition in this region.

%%%%%%%%%%%%%%%%%%%%%%%%%%%%%%%%%%%%%%%%%%%%%%%%%%%%%%%%%%%
%%%%%%%%%%%%%%%%%%%%%%%%%%%%%%%%%%%%%%%%%%%%%%%%%%%%%%%%%%%
\section{Phenomenology}\label{sec:pheno}
%%%%%%%%%%%%%%%%%%%%%%%%%%%%%%%%%%%%%%%%%%%%%%%%%%%%%%%%%%%
%%%%%%%%%%%%%%%%%%%%%%%%%%%%%%%%%%%%%%%%%%%%%%%%%%%%%%%%%%%

\subsection{The dilaton}
\begin{figure}
\centering
\includegraphics[width=6in]{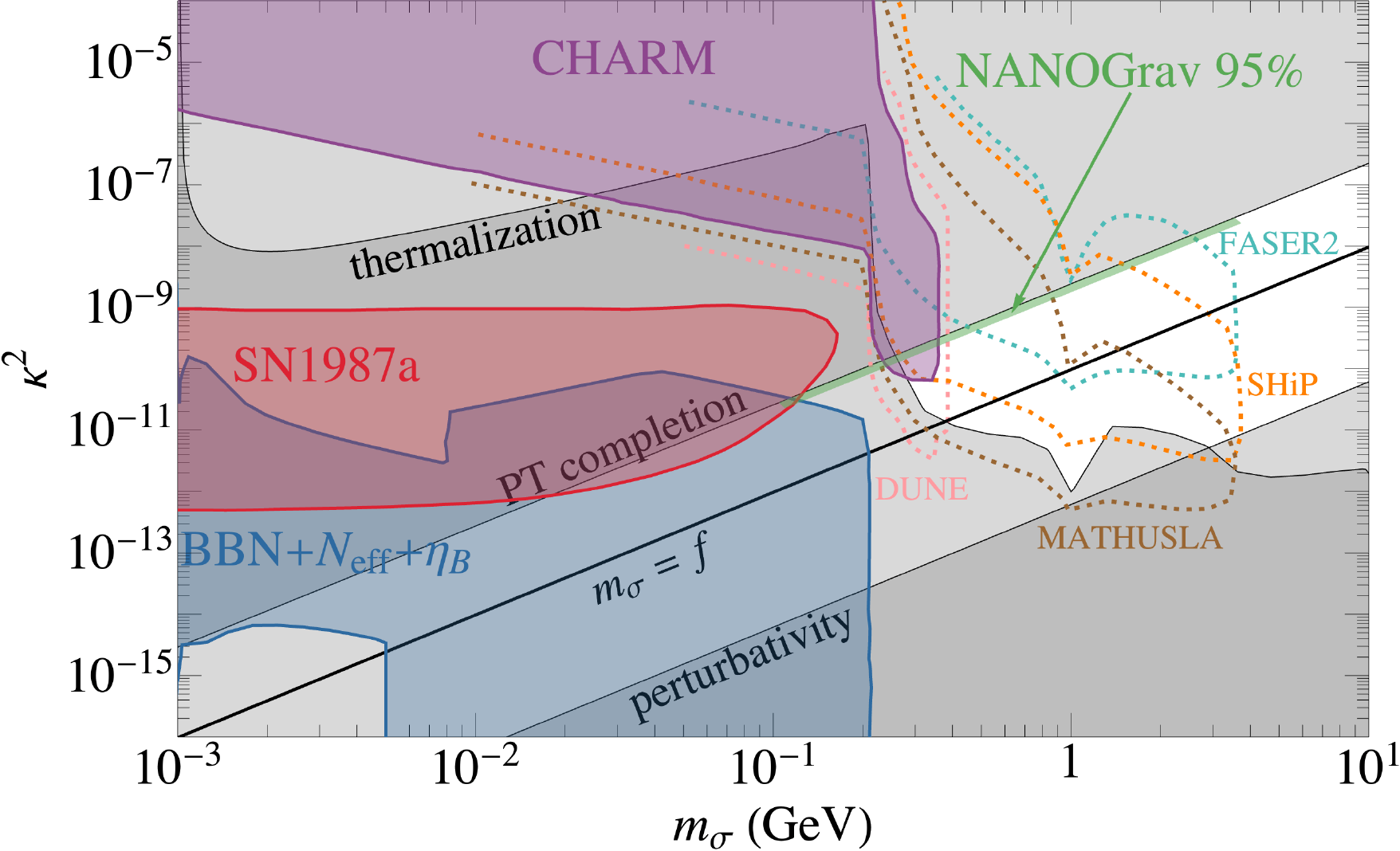}
    \caption{Exclusion plot for the dilaton as a function of its mass $m_{\sigma}$ and coupling strength $\kappa^{2}$, fixing $\Lambda = 5$~TeV, $\lambda = 1$, and $\alpha=0.15$. We include a contour for $f = m_\sigma$ \textit{(black line)}. The gray shaded regions indicate the theoretical bounds that the dilaton thermalize with the SM \textit{(``thermalization'')}, that the dilaton effective theory is valid \textit{(``perturbativity'')}, and that the phase transition completes \textit{(``PT completion'')}. We show experimental bounds, adapted from~\cite{Flacke:2016szy}, from late decays \textit{(blue)}~\cite{Planck:2015fie,Poulin:2015opa}, from supernova cooling \textit{(red)}~\cite{Krnjaic:2015mbs}, and from the CHARM experiment \textit{(purple)}~\cite{CHARM:1985anb}, as well as projected sensitivities for DUNE \textit{(pink line)}~\cite{Berryman:2019dme}, FASER2 \textit{(turquoise line)}~\cite{Feng:2022inv}, MATHUSLA \textit{(brown line)}~\cite{Curtin:2018mvb}, and SHiP \textit{(orange line)}~\cite{Alekhin:2015byh}. We also show the parameter space which reproduces the NANOGrav signal at the 95\% credible level \textit{(green)}. }
\label{fig:DilatonExclusionPlot}
\end{figure}

There is a rich phenomenology associated with the dilaton in our model. From Eq.~\eqref{eq:effL}, we see that the dilaton has the same couplings to the SM fermions and gauge bosons as the Higgs, but rescaled by a factor
\begin{equation}
    \kappa = \frac{v f}{\Lambda^2} .
\end{equation}
In Fig.~\ref{fig:DilatonExclusionPlot} we show theoretical and experimental constraints on the dilaton in the $(m_\sigma, \kappa^2)$ plane, fixing the cutoff scale $\Lambda = 5$~TeV and the dilaton potential parameters $\lambda = 1$ and $\alpha = 0.15$. The latter are only relevant for phase transition calculations. We also present a contour for $m_\sigma = f$ in Fig.~\ref{fig:DilatonExclusionPlot} and indicate where the conformal phase transition yields a stochastic gravitational wave signal consistent with the NANOGrav result (see section~\ref{sec:pt}).

%%%%%%%%%%%%%%%%%%%%%%%%%%%%%%%%%%%%%
%%%%%%%%%%%%%%%%%%%%%%%%%%%%%%%%%%%%

There are three important theoretical bounds we consider, which we depict by gray shaded regions in Fig.~\ref{fig:DilatonExclusionPlot}. First, for the dilaton effective theory to be valid we must have $m_\sigma < 4\pi f$. Second, there is a lower bound on dilaton mass imposed by the requirement that the conformal phase transition can complete, as discussed in Section~\ref{sec:pt}. Finally, it is crucial that the dilaton is in thermal contact with the SM following the conformal phase transition, or the dark sector can never thermalize with the SM. We crudely estimate this bound by computing the decay width $\Gamma$ of the dilaton, following~\cite{Winkler:2018qyg}, and requiring that $\Gamma > H$ at $T = m_\sigma$. From Fig.~\ref{fig:DilatonExclusionPlot} we see that these three constraints are quite powerful, together implying that the dilaton must be heavier than $\sim 0.1$~GeV and have a coupling strength $\kappa^2 \gtrsim 10^{-12}$.

The experimental bounds are similar to a scalar singlet mixing with the Higgs through an angle $\sin \theta = \kappa$. In Fig.~\ref{fig:DilatonExclusionPlot} we present the following constraints adapted from~\cite{Flacke:2016szy}: cosmological bounds on late decays of the dilaton to SM particles\footnote{Late decays can alter the effective number of relativistic neutrino species $N_{\rm eff}$~\cite{Planck:2015fie}, the baryon-to-photon ratio $\eta_B$~\cite{Poulin:2015opa}, and the light element abundances predicted by big bang nucleosynthesis (BBN)~\cite{Poulin:2015opa}.}~\cite{Planck:2015fie,Poulin:2015opa}, astrophysical bounds from supernova cooling~\cite{Krnjaic:2015mbs}, and bounds from the beam dump experiment CHARM~\cite{CHARM:1985anb}. Only the last of these excludes parameter space not already ruled out by the theoretical constraints discussed above. There also exist constraints from searches for rare meson decays, but these are too weak to exclude any relevant parameter space, so we omit it from Fig.~\ref{fig:DilatonExclusionPlot} for clarity.

We also show projections for future experimental bounds in Fig.~\ref{fig:DilatonExclusionPlot}. Experiments searching for light, weakly-coupled particles such as FASER2~\cite{Feng:2022inv}, MATHUSLA~\cite{Curtin:2018mvb}, and SHiP~\cite{Alekhin:2015byh} (and also DUNE~\cite{Berryman:2019dme}) will probe much of the remaining parameter space around $m_\sigma \sim 1$~GeV. Notably, this includes nearly all of the region favored by the NANOGrav signal. 

\subsection{DM phenomenology}
\begin{figure}
\centering
\includegraphics[width=6in]{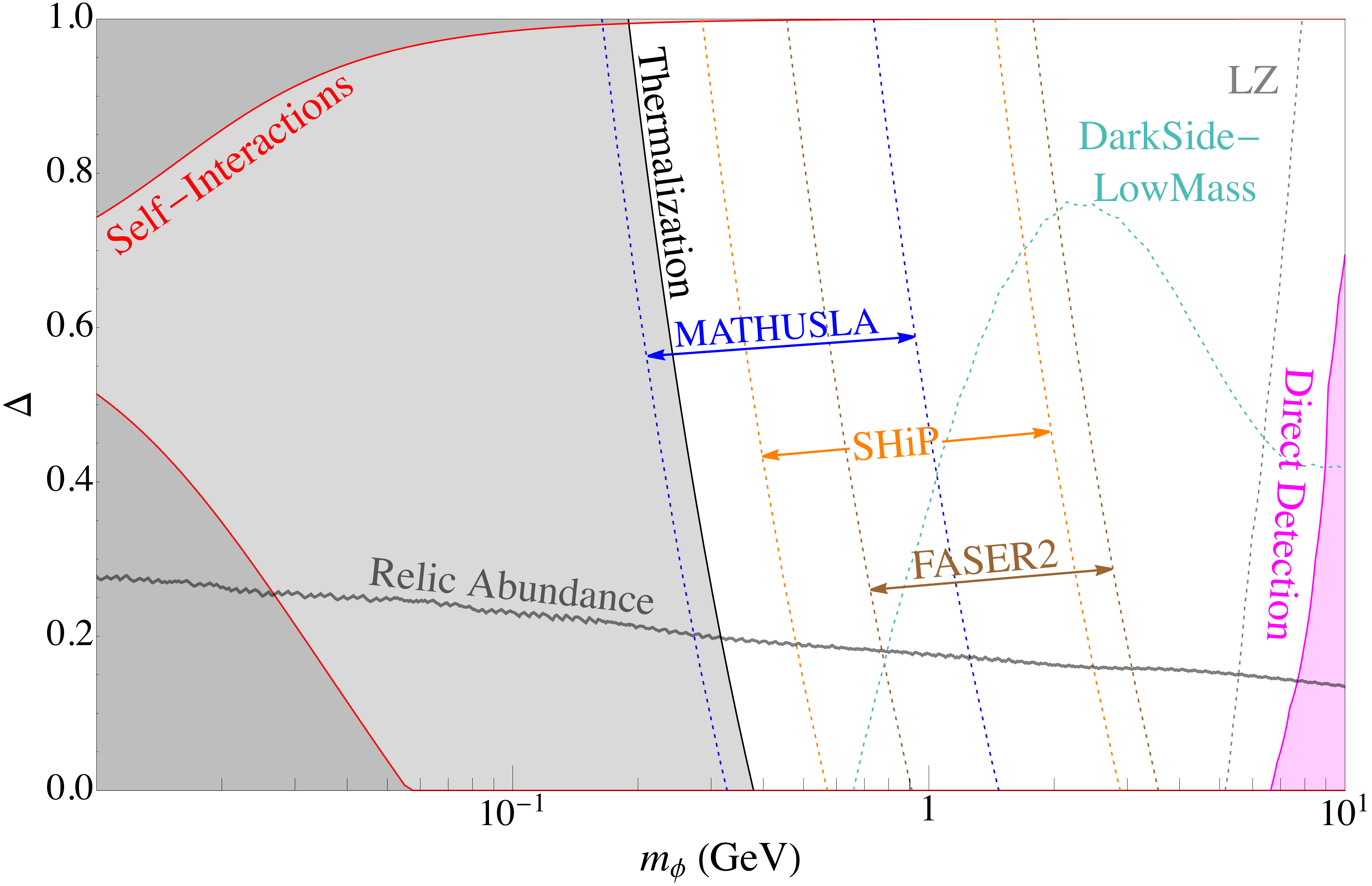}
\includegraphics[width=6in]{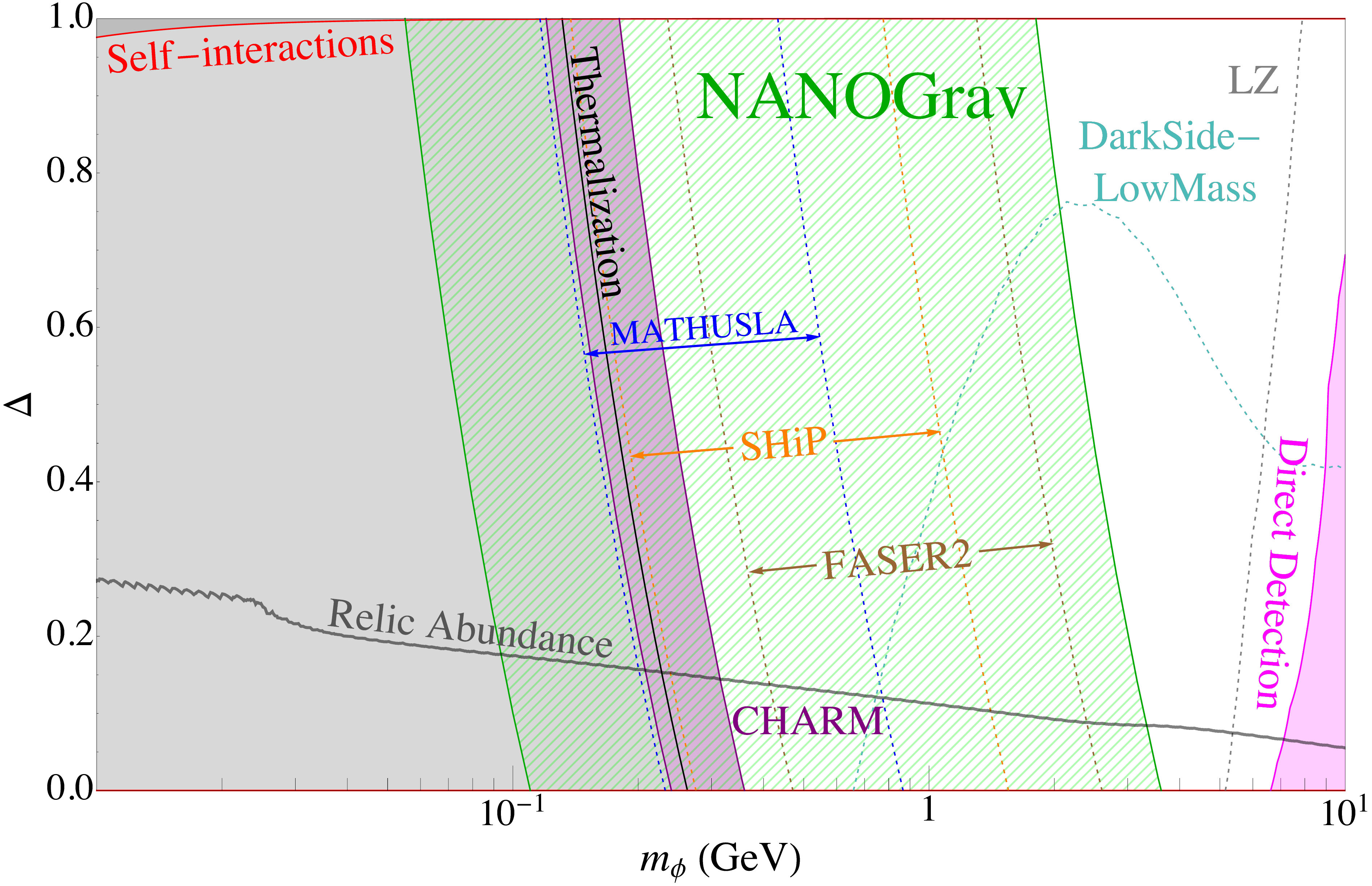}
    \caption{Exclusion plot as a function of DM mass $m_{\phi}$ and mass splitting $\Delta$ for $f/m_{\sigma} = 1$ \textit{(top)} and $f/m_{\sigma} = 4.5$ \textit{(bottom)}, fixing $\Lambda = 5$~TeV, $\lambda = 1$, and $\alpha=0.15$. The gray lines reproduce the observed DM relic abundance. We include applicable bounds on the dilaton (see Fig.~\ref{fig:DilatonExclusionPlot} for details). We present bounds on the DM from self-interactions \textit{(red)}~\cite{Harvey:2015hha,Tulin:2017ara,ParticleDataGroup:2022pth} and direct detection \textit{(magenta)}~\cite{PandaX-4T:2021bab,LZ:2022lsv,XENON:2023cxc}, as well as projections, adapted from~\cite{Akerib:2022ort}, for LZ with 15 ton-years exposure \textit{(gray line)}~\cite{LZ:2018qzl} and for DarkSide-LowMass with 1 ton-year exposure \textit{(turquoise line)}~\cite{GlobalArgonDarkMatter:2022ppc}.}
\label{fig:DMExclusionPlot}
\end{figure}

Having studied the phenomenology of the dilaton $\sigma$, we now turn to the DM $\phi$. In Fig.~\ref{fig:DMExclusionPlot} we show contours in the $(m_\phi, \Delta)$ plane that yield the observed DM relic abundance for $f/m_\sigma = 1$ and $f/m_\sigma = 4.5$ (still fixing $\Lambda = 5$~TeV), as well as all the relevant constraints. The choice of $f/m_\sigma = 4.5$ allows for a stochastic gravitational wave background consistent with the NANOGrav signal for a range of dilaton masses, which we indicate by a green band in the lower panel of Fig.~\ref{fig:DMExclusionPlot}. We additionally include theoretical and experimental bounds on the dilaton, which were discussed above, and bounds on $\phi$ from DM self-interactions and direct detection, which we discuss below.

%%%%%%%%%%%%%%%%%%%%%%%%%%%%%%%%%%%%%%%%%%%%%%%%%%%%%%%%%
%%%%%%%%%%%%%%%%%%%%%%%%%%%%%%%%%%%%%%%%%%%%%%%%%%%%%%%%%

Evidence from merging galaxy clusters places a limit on DM self-interactions $\sigma_{\rm SI}/m_\phi \lesssim 0.5{\rm ~cm^2/g} \approx 2000 {\rm ~GeV^{-3}}$~\cite{Harvey:2015hha,Tulin:2017ara,ParticleDataGroup:2022pth}. Self-scattering $\phi\phi \rightarrow \phi\phi$ can be mediated by dilaton exchange or proceed through the contact interaction $\lambda_\phi \phi^4 / 4!$. In the nonrelativistic limit, we find a transfer cross section of 
\begin{equation}
    \sigma_{\rm SI} = \frac{m_\phi^2}{32\pi f^4} \left[ \frac{8}{(1+\Delta)^2} + \frac{4}{(\Delta+3)(\Delta-1)} - \lambda_\phi \right]^2 .
\end{equation}
The presence of the contact term can either increase or decrease $\sigma_{\rm SI}$ due to interference with the dilaton exchange diagrams. For concreteness we set $\lambda_\phi = 0$ to compute explicit bounds in Fig.~\ref{fig:DMExclusionPlot}.

The most stringent experimental bounds on $\phi$ arise from direct detection constraints on the DM-nucleon cross section. Due to the QCD trace anomaly, the dilaton effectively couples to the total mass of hadrons~\cite{Damour:2010rp,Kaplan:2000hh}. Furthermore we can integrate out the dilaton at small momentum transfer, leading to the effective DM-nucleon interaction
\begin{equation}
    \frac{1}{2} \phi^2 \overline{n} n \frac{2 m_n}{(1+\Delta)^2 \Lambda^2} 
\end{equation}
where $m_n$ is the nucleon mass. Using this we find a DM-nucleon cross section of
\begin{equation}
    \sigma_{\phi N} = \frac{m_n^4}{\pi (1+\Delta)^4 \Lambda^4 (m_\phi + m_n)^2 } .
\end{equation}
Comparing this to the leading bounds from PandaX-4T~\cite{PandaX-4T:2021bab}, XENONnT~\cite{LZ:2022lsv}, and LZ~\cite{XENON:2023cxc} allows one to place a lower bound on the mass splitting $\Delta$ as a function of $m_\phi$. From Fig.~\ref{fig:DMExclusionPlot} we see this effectively rules out DM with mass larger than $7$~GeV.

We also include projections for future direct detection bounds in Fig.~\ref{fig:DMExclusionPlot} adapted from~\cite{Akerib:2022ort}. In particular, the DarkSide-LowMass experiment~\cite{GlobalArgonDarkMatter:2022ppc} would be sensitive to DM masses above a GeV. This complements the bounds from future searches for long-lived particles (like FASER2, MATHUSLA, and SHiP): all together these experiments will probe the entire remaining parameter space of our model.

\subsection{Other constraints}

The DM may annihilate to SM fermions through the dilaton portal, but the resulting cross section is well below observational limits. In the limit where the fermion is much lighter than the DM, the thermally averaged annihilation cross section is
\begin{equation}
    \langle \sigma v (\phi \phi \rightarrow f \overline{f}) \rangle \sim 10^{-36} {\rm ~cm^3/s} \left((1-\Delta)(3+\Delta)\right)^{-2} \left( \frac{m_f}{0.5{\rm ~MeV}} \right)^2 \left( \frac{1 {\rm ~TeV}}{\Lambda} \right)^4
\end{equation}
where $m_f$ is the fermion mass. For all annihilation channels, this cross section is several orders of magnitude below the experimental constraints on electromagnetic energy injection in the early universe~\cite{Slatyer:2015jla}. In principle the cross section gets a Sommerfeld enhancement, but the effect is far too small for it to approach the experimental bounds. To see this, observe that the dilaton mediates a force between DM particles with an effective ``fine-structure constant'' of $\alpha_{\rm eff} = m_\phi^2 / (4 \pi f^2)$. Sommerfeld enhancement is only a large effect when the quantity
\begin{equation}
    \epsilon = \frac{m_\sigma}{\alpha_{\rm eff} m_\phi} = 4\pi (1+\Delta)^3 \frac{f^2}{m_\sigma^2}
\end{equation}
is much smaller than $1$~\cite{Arkani-Hamed:2008hhe,Agashe:2009ja}. This is only satisfied for a relatively heavy dilaton. Numerically, for $f = m_\sigma$ and a DM velocity of $0.5 \times 10^{-3}$, we find an enhancement of $2\%$ to $17\%$, depending on the value of $\Delta$.

Lastly, the Higgs can decay to KK modes of the dilaton through a brane-localized interaction with the Goldberger--Wise scalar. Although we did not write such an interaction in the 5D Lagrangian in section~\ref{sec:model}, it is allowed by the symmetries of our model. These decays contribute to the invisible width of the Higgs, which is constrained by collider experiments. We can estimate the decay width to KK modes as follows. We expect that the amplitude for the process $h \rightarrow {\rm KK}+{\rm KK}$ is given by $\Lambda$ times a wavefunction suppression factor. Since KK wavefunctions are given by Bessel functions which asymptotically grow like $z^{3/2}$ (with $z$ the coordinate along the extra dimension as in Eq.~\eqref{eq:5Dmetric}), we expect a wavefunction suppression of $(f / \Lambda)^3$ for a final state with two KK modes. Thus the decay width should scale like
\begin{equation}\label{eq:higgstoKK}
    \Gamma (h \rightarrow {\rm KK} + {\rm KK}) \sim \frac{\Lambda^2}{8\pi m_h} \left( \frac{f}{\Lambda} \right)^6 .
\end{equation}
The number of KK modes lighter than the Higgs is of order $m_h / f$, so the total number of decay channels --- each with a width as in Eq.~\eqref{eq:higgstoKK} --- is of order $m_h^2 / f^2$. Thus, the total invisible decay width is
\begin{equation}
    \Gamma (h \rightarrow {\rm invisible}) \sim \frac{m_h}{8 \pi} \left( \frac{f}{\Lambda} \right)^4 .
\end{equation}
ATLAS constrains the Higgs invisible branching ratio to be less than $0.11$~\cite{ATLAS:2023tkt}, which implies a very weak bound $\Lambda / f \gtrsim 10$. This is clearly satisfied for our case of interest with $f \sim $~GeV and $\Lambda \sim$~TeV.

%%%%%%%%%%%%%%%%%%%%%%%%%%%%%%%%%%%%%%%%%%%%%%%%%%%%%%%%%%%
%%%%%%%%%%%%%%%%%%%%%%%%%%%%%%%%%%%%%%%%%%%%%%%%%%%%%%%%%%%
\section{Discussion and conclusions}\label{sec:conclusions}
%%%%%%%%%%%%%%%%%%%%%%%%%%%%%%%%%%%%%%%%%%%%%%%%%%%%%%%%%%%
%%%%%%%%%%%%%%%%%%%%%%%%%%%%%%%%%%%%%%%%%%%%%%%%%%%%%%%%%%%
In this paper we have explored a model of forbidden conformal DM. General considerations of GeV-scale, thermal relic DM from a conformal sector naturally led us to consider the regime where the dilaton $\sigma$ is heavier than the DM $\phi$, such that the forbidden channel $\phi \phi \rightarrow \sigma \sigma$ determined the relic abundance. We studied this model in a 5D dual implementation, consisting of a warped extra dimension where the DM propagates on the IR brane and the SM propagates on the UV brane. We studied the conformal phase transition in our model and found that for a range of dilaton masses around $0.1$--$2$~GeV, the phase transition can source a nHz-scale stochastic gravitational wave background consistent with that observed at NANOGrav. Note that this setup can be easily UV-completed into a composite Higgs model by extending it to a three-brane model, and in this case we would have additional phase transition that would show up as a double peak signature that can be probed by for the future space-based gravitational wave experiments. 

Theoretical and experimental bounds pointed to dark sector masses in the range $0.1$--$10$~GeV. Imposing the requirements that the dark sector thermalizes with the SM, that the conformal phase transition completes, and that the dilaton effective theory is valid led to a lower bound on the dilaton mass of about $0.1$~GeV; meanwhile, direct detection bounds constrained the DM mass to be less than $10$~GeV. The viable parameter space below a few GeV will be probed by experiments searching for light, weakly-coupled particles like FASER2, MATHUSLA, and SHiP. Future direct detection experiments specialized for low mass WIMPs, in particular  DarkSide-LowMass, will be sensitive to the remaining parameter space up to $10$~GeV.

To the best of our knowledge, our work represents the first extensive study of light thermal relic DM which is a composite of a CFT. We have focused on forbidden DM here, but there is probably a wealth of model-building opportunities to be found in exploring other thermal mechanisms for generating light conformal DM. The detection of keV--GeV DM is a very active area of research with exciting recent developments~\cite{Kahn:2021ttr}. Thus, it would be particularly interesting to identify a mechanism that allows for conformal DM lighter than the $0.1$~GeV lower bound we found in our model.

\acknowledgments 
We would like to thank Alexander Pukhov for helping us deal with the small DM mass in \texttt{micrOMEGAs}.
SF and AI are supported in part by the NSF grant PHY-2014071. AI is also supported in part by NSERC, funding reference number 557763. The research activities of SL and YL are supported by the Samsung Science Technology Foundation under Project Number SSTF-BA2201-06, and also by the National Research Foundation of Korea (NRF) grant funded by the Korea government (MEST) (No. NRF-2021R1A2C1005615).

\bibliographystyle{JHEP}
\bibliography{referencesv2}

\end{document}